\documentstyle[12pt]{article}
\begin{document}

\title{
\begin{flushright}
{\normalsize VPI-IPPAP-99-04}\\ \ \\
\end{flushright}
Constraining SUSY Models with Spontaneous CP-Violation via
$B\rightarrow \psi K_{s}$}
\author{ Oleg Lebedev\\
     \it{Virginia Polytechnic Insitute and State University}\\
     \it{Department of Physics}\\
     \it{Blacksburg, Virginia 24061-0435}}
\maketitle
\thispagestyle{empty}
\begin{abstract}
We study CP-violating effects in $B\rightarrow \psi K_{s}$ decay within
minimal
supersymmetric models with spontaneous CP-violation.
We find that the CP-asymmetry predicted by the Standard Model in this decay,
$\sin2\beta\geq 0.4$, cannot be accommodated in these models without
violating the bound on the neutron
electric dipole moment. This result holds for NMSSM-like models with
an arbitrary number of sterile superfields. Further implications of the
scenario are discussed.  \\ \ \\
\\ \ \\
\end{abstract}
\newpage
\section{Introduction}
The origin of CP-violation is one of the most profound
problems in particle physics.
In the Standard Model, all observable CP-violating
effects in the kaon system can be successfully explained via the
Cabibbo-Kobayashi-Maskawa (CKM) mechanism [1].
However, the physical principles
lying behind CP-violation are still not understood.

One of the more elegant approaches to the problem of CP-violation
is based on the possibility of spontaneous T-breaking in multi Higgs
doublet systems [2]. Supersymmetric models can provide such systems
and thus are a natural setting to implement this idea. Spontaneous
CP-violation (SCPV)
in susy models has drawn considerable attention [5-10] due to
the following attractive features: \\
$\bullet$ CP-phases become dynamical variables,\\
$\bullet$ CP-symmetry is restored at high energies,\\
$\bullet$ it allows to avoid excessive CP-violation inherent in susy models.

In this letter, we consider the implications of 
SCPV in minimal susy models for CP-asymmetries in
$B\rightarrow \psi K_{s}$ decays.
It is well known that the Standard Model predicts  large CP-violation in
these decays, namely $\sin2\beta\geq 0.4$ where $\beta$ is the one of
the angles of the unitarity triangle [21]. Currently this CP-asymmetry 
is being studied experimentally by the CDF collaboration. Even though
the statistics
does not allow to make definite statements about the validity of the SM
predictions at the moment, large CP-violation in this decay has been hinted.
The purpose of this paper is to determine whether a large
CP-asymmetry, $\sin2\beta\geq 0.4$,
  can be explained in susy
models with spontaneously broken CP.

\section{CP-Asymmetry in $B\rightarrow \psi K_{s}$ Decay
and Spontaneous CP-Violation}
In this section we will consider minimal susy models with
spontaneously broken CP and
obtain the lower bound on the CP-violating phase
as dictated by $\sin2\beta\geq 0.4$ .

It has been shown that the Next-to-Minimal Supersymmetric Standard Model
(NMSSM) is the simplest susy model which allows spontaneous
CP-violation while being consistent with the experimental bound on the
lightest Higgs mass [8]. In the most general version of the NMSSM
with the superpotential
\begin{eqnarray}
&& W= \lambda \hat N\hat H_1\hat H_2 -{k\over3}\hat N^3 -r\hat N
+\mu \hat H_1 \hat H_2 + W_{fermion}\;,
\end{eqnarray}
SCPV can occur already at the tree level thereby avoiding the Georgi-Pais
theorem [6]. Note that even though SCPV in the MSSM
is allowed theoretically [5], such a scenario is ruled out by the LEP
constraints on the axion mass [7].

We will assume that $all$ CP-violating effects result from the complex
Higgs VEV's
\begin{eqnarray}
&&\langle H_1^0 \rangle =v_1,\;\langle H_2^0 \rangle=v_2 e^{i\rho},
\langle N \rangle=n e^{i\xi}\;. 
\end{eqnarray}
The relevant interactions{\footnote{The complete list of interactions
can be found in [12].}} for one generation of fermions
(after spontaneous $SU(2)\times U(1)$ symmetry
breaking) can be written as follows [9,12]:
\begin{eqnarray}
 {\mathcal{L}}&=&-h_{u} H_2^0 {\bar u_{R}} u_{L}- h_{d} H_1^0
   {\bar d_{R}} d_{L} -g {\overline {\tilde W^{c}}} P_{L} d\;\tilde u^{*}_{L}
   +h_{d} \bar d P_{L} \tilde H^{c} \tilde u_{L} \nonumber\\
   &+&  h_{u} {\overline {\tilde H^{c}}} P_{L} d\; \tilde u_{R}^{*}+h.c.\;,     \\
 {\mathcal{L}}_{mix}&=&-g(\;v_1 {\overline {\tilde H}} P_{L} \tilde W +
   v_2 e^{-i\rho} {\overline {\tilde W}} P_{L} \tilde H\;)+
   e^{-i\kappa_{u}} h_{u} {m_{LR}^{(u)}}^2 {\tilde u_{R}^{*}} \tilde u_{L}
   \nonumber \\
   &+& e^{-i\kappa_{d}} h_{d} {m_{LR}^{(d)}}^2 {\tilde d_{R}^{*}} \tilde d_{L}
   +h.c.\;,
   \end{eqnarray}
where $h_{u,d}$ denote the Yukawa couplings and $\kappa_{u,d}$ are certain
functions of the Higgs VEV phases $\rho$ and $\xi$. In what follows, we
assume, for the sake of definiteness, that $| \kappa_{u}|=| \kappa_{d} |=
| \kappa |$, ${m_{LR}^{(u)}}^2={m_{LR}^{(d)}}^2=m_{LR}^2$, and that
$\kappa$ and $m_{LR}^2$ are generation independent. Equation (4) represents
the wino-higgsino and left-right squark mixings, with the former being
responsible for the formation of the mass eigenstates - charginos.
However, following Ref. [9],
we will prefer the ``weak" (wino-higgsino) basis over the mass (chargino) one.

Information about the angle $\beta$ of the unitarity triangle was
extracted from the decay rate evolution
$$\Gamma (B^{0} [\bar B^{0}](t)\rightarrow \psi K_{s}) \;\propto e^{-\Gamma t} \biggl(
1-[+] \sin 2 \beta \;\sin \Delta m t \biggr)\;,$$
where $\Delta m$ is the $B_{L}-B_{H}$ mass difference. 
On the other hand, the angles of the
unitarity triangle can be expressed in terms of the mixing ($\phi_{M}$) and
decay ($\phi_{D}$) phases which enter the $B-\bar B$ mixing
and $b\rightarrow q\bar qQ$ decay diagrams. For the process under
consideration,
\begin{eqnarray}
&&\sin2\beta =\sin(2\phi_{D}+\phi_{M})\;.
\end{eqnarray}
This relation is theoretically clean since it does not involve hadronic
uncertainties and can serve as a sensitive probe  for physics
beyond the Standard Model.
At the present time, the CKM entries are not known precisely enough
to make a definite prediction for $\beta$. However, it is known
that $\sin 2\beta$ must fall between 0.4 and 0.9 in order for the
CKM model to be consistent [21].
In our model, we will impose this condition
 together with Eq.(5) to obtain the
lower bound on the CP-violating phases appearing in the decay and mixing
diagrams.

Let us now proceed to calculating the CP-violating effects in $B-\bar B$
mixing. Figure 1 displays what we believe to be the most important
contributions to the real part of $B-\bar B$ mixing. These include the
Standard Model box and wino superbox diagrams.
It has been argued [9] that all significant CP-violating effects
result from complex phases in the propagators of the superparticles.
For the $K-\bar K$ system, one loop diagrams involving complex phases in
the wino-higgsino and left-right squark mixings can lead to the correct
values of $\epsilon$ and $\epsilon'$ [9].
The analogous $B-\bar B$ diagrams are shown in Figure 2 (in the case of
$K-\bar K$ mixing, the diagram in Fig. 2b is super-CKM suppressed and
can be neglected). Besides the above mentioned contributions, there is
a number of other contributions to $B-\bar B$ mixing
which can be classified as follows:\\
1. Higgs boxes,\\
2. gluino superboxes,\\
3. neutralino superboxes.\\
The gluino contribution can
be neglected since the gluino is likely to be very heavy:
$m_{\tilde g} \geq 310\; GeV$ [13]. Further, the neutralino 
analogs
of Figs.2a,b have to involve at least two powers of $h_{b}$ and, thus,
are suppressed by $(m_{b}/m_{W})^2$.
The same argument applies to the charged Higgs contribution [10].
On the other hand, the neutralino
and Higgs
contributions to $Re\;(B-\bar B)$ can be significant. 
However, they interfere with the SM contribution constructively [14]
and can only reduce the $B-\bar B$ mixing weak phase. Since
our purpose is to determine the $lower$ bound on this phase as
dictated by $\sin 2\beta \geq 0.4$, we can safely omit these
corrections.

Therefore, for our purposes it is sufficient to retain the SM and
chargino contributions only. Furthermore, note that we may concentrate on the
$(V-A)\times (V-A)$ interaction solely since 4-fermion chargino-generated
interactions involving the right handed quarks are suppressed by powers of
$m_{b}/m_{W}$. As a consequence, hadronic uncertainties and QCD corrections
will not affect our results since those will cancel in the expression for
the phase. The resulting 4-fermion interaction can be expressed as
\begin{eqnarray}
{\mathcal{O}}_{\Delta B=2} &=& (\;k_{SM} + k_{SUSY}+ e^{i\delta}l_2 +
ze^{-i\delta}l_2 +e^{2i\delta}l_4 + z^2 e^{-2i\delta}l_4\;) \nonumber\\
&\times&\bar d\gamma^{\mu} P_{L} b\;\bar d\gamma^{\mu} P_{L} b \;,
\end{eqnarray}
where $k_{SM}$ and $k_{SUSY}$ are real couplings induced by the diagrams
in Fig.1a and Fig.1b, respectively. The CP-violating couplings
$e^{i\delta}l_2$ and $e^{2i\delta}l_4$ result from the diagrams with
two and four complex mixings shown in Figs.2a and 2b, respectively ($l_{2,4}$
are defined to be real). It is important to note that along with the
diagrams explicitly shown in Fig.2, there are also ``cross" diagrams in which
the positions of $\tilde t_{L}$ and $\tilde t_{R}$ are interchanged. Such
graphs contribute with opposite phase and may seem to lead to a complete
cancellation of the imaginary part of the coupling. However, this
cancellation is only partial [9] owing to the fact that the higgsino
vertex is, generally speaking, different from that of gaugino. Such partial
cancellation is accounted for by a variable factor
$z$ ($0\leq z \leq 1$). 

The Standard Model contribution is given by [15]{\footnote{
We do not show the QCD correction factor explicitly.}}
\begin{eqnarray}
&& k_{SM}={G_{F}^2\over 16 \pi^2} m_{t}^2 H(x_{t})\;(V_{tb}V_{td})^2\;,
\end{eqnarray}
with $H(x)$ being the Inami-Lim function.
To estimate the superbox contributions, we may treat the gaugino and higgsino
as particles of mass $m_{\tilde W}$ with (perturbative)
mixing given by Eq.(4).
In this approximation, the chargino propagator with a mixing insertion (Fig.2)
is proportional to $g(v_1+v_2 e^{-i\rho}) {m_{\tilde W} \not k\over
(k^2-m_{\tilde W}^2)^2}$. The resulting 4-fermion couplings  are given
by
\begin{eqnarray}
k_{SUSY} &=& {g^4 \zeta^2 \over 128 \pi^2}\;{1\over m_{\tilde q}^2}\;
             ({\tilde V_{tb}} {\tilde V_{td}})^2\;
             {1\over (y-1)^2}\biggl[ y+1-{2y\over y-1}\ln y \biggr]\;,
             \\             
e^{i\delta}l_2  &=& {g^4 h_{t}^2 \zeta \over 64 \pi^2}\;
   {m_{LR}^2 e^{-i\kappa}\;(v_1+v_2 e^{-i\rho}) \over m_{\tilde q}^5} \;
   ({\tilde V_{tb}} {\tilde V_{td}})^2 \nonumber\\
   &\times& {\sqrt{y} \over(y-1)^5}\biggl[ 3-3y^2+(1+4y+y^2)\ln y \biggr]
   \;,\\
e^{2i\delta}l_4  &=& {g^4 h_{t}^4 \over 384 \pi^2}\;
   {m_{LR}^4 e^{-2i\kappa}\;(v_1+v_2 e^{-i\rho})^2 \over m_{\tilde q}^8} \;
   ({\tilde V_{tb}} {\tilde V_{td}})^2 \nonumber\\
   &\times& {1\over(y-1)^7}\biggl[ -1-28y+28y^3+y^4-12y(1+3y+y^2)\ln y \biggr]
   \;,
\end{eqnarray}
where $y=m_{\tilde W}^2/m_{\tilde q}^2\;$; $m_{\tilde q}$ and $m_{\tilde W}$
denote the top
squark and the chargino masses,
respectively, and
${\tilde V}$ is the squark analog of the CKM matrix.
In these considerations, the top squark contribution is believed to play
the most important role. The influence of the $c$- and $u$-squarks is
taken into account through a variable super-GIM cancellation factor $\zeta$
($0\leq \zeta \leq 1$).
Such a factor is associated with every squark line on which summation
over all the up-squarks  takes place. Since masses of the top and
$c,u$-squarks are expected to be very different due to the large top
Yukawa coupling (as motivated by SUGRA),
the natural value of $\zeta$ would be of order unity
($\zeta$ is defined to vanish in the limit of degenerate squarks).

To derive the lower bound on the phase $\delta$, we may replace
$v_1+v_2 e^{-i\rho}$ in Eqs.(9) and (10)
by its ``maximal" value $\sqrt{2} v e^{-i\rho}$
with $v$ defined in the usual way: $v=\sqrt{v_1^2+v_2^2} \approx 174 GeV$.
Apparently, $\delta \leq |\rho|+|\kappa|\;.$

Let us now determine the weak phases $\phi_{M}$ and $\phi_{D}$.
It follows from Eq.(6) that
\begin{eqnarray}
&& \tan \phi_{M} = {l_2 (1-z)\;\sin\delta +l_4 (1-z^2)\;\sin2\delta \over
k_{SM} +k_{SUSY}+l_2 (1+z)\;\cos\delta +l_4 (1+z^2)\;\cos2\delta}\;.
\end{eqnarray}
On the other hand, the decay phase $\phi_{D}$ is negligibly small [10].
Indeed, in our model, the only source of CP-violation in the process
$b\rightarrow c\bar c s$ is the superpenguin diagram with the charginos
and squarks in the
loop{\footnote{Another possible contributor, Higgs-mediated tree level decay,
is suppressed by the quark masses.}}.
However, this diagram is greatly suppressed
as compared to its CP-conserving counterpart, $W$ mediated tree level
decay, due to the loop factors and heavy squark propagators.
Also, unlike for the kaon decays, there is no $\Delta I=1/2$
enhancement of the $(V-A)\times (V+A)$ interactions. As a result,
the weak decay phases can be neglected. This also means that
direct CP-violation in our model is negligibly small as compared
to that in the Standard Model (unless there is no tree level decay mode).

According to Eq.(5), the angle $\beta$ is determined by
\begin{eqnarray}
&& \sin2\beta= \sin\phi_{M}\;.
\end{eqnarray}
Then the experimental bound $\sin2\beta \geq 0.4$ can be translated into
\begin{eqnarray}
&& \tan\phi_{M} \geq 0.44\;.
\end{eqnarray}
This, in turn, leads to a lower bound on the phase $\delta$ which
can be obtained numerically from Eq.(11).
Note that Eq.(11) is free of hadronic uncertanties and QCD
radiative corrections.

\section{Numerical Results}
In this section we will discuss implications of Eq.(13)
and  its compatibility with the upper bound on the NEDM.

It is well known that the tight experimental bound on the NEDM
imposes stringent constraints on the CP-violating phases which appear
in extensions of the Standard Model. In our model, the largest
contributions to the NEDM are shown in Fig.3. Barring accidental
cancellations, one can constrain the CP-phases entering the
higgsino-gaugino and squark left-right mixings via the
chargino (Fig.3b) and gluino (Fig.3a) contributions
to the NEDM, respectively. Consequently, the phase $\delta$ appearing
in the $B-\bar B$ mixing becomes bounded due to
\begin{eqnarray}
&&\sin\delta\leq |\sin\kappa | + |\sin\rho |\;.
\end{eqnarray}
This requires $\delta$ to be of order $10^{-2}$-$10^{-1}$ [9,16].

On the other hand,
Eq.(13) imposes a lower bound on $\delta$. 
As seen from Eqs.(8)-(10),
this bound depends strongly on the super-CKM matrix.
We consider the following possibilities:\\
1. the super-CKM matrix duplicates the CKM one,
${\tilde V_{td}}\approx V_{td}$;\\
2. the super-CKM mixing is enhanced,
${\tilde V_{td}}\approx V_{td}/\sin\theta_{C}$;\\
3. the super-CKM mixing is doubly enhanced,
${\tilde V_{td}}\approx V_{td}/(\sin\theta_{C})^2$.\\
In all of these cases we assume $\tilde V_{tb}\sim 1$. The first
possibility implies that the supersymmetric contribution to $B-\bar B$
mixing is suppressed as compared to the CP-conserving Standard Model box
diagram. As a result, we find that the constraint
$\tan \phi_{M} \geq 0.44$ cannot be satisfied for any $\delta$ even assuming
light ($100 \;GeV$) squarks. For the same reason, we are bound to consider
only light ($100\;GeV$) chargino and maximal left-right
squark mixing, $m_{LR}=m_{\tilde q}$.
On the other hand, the third option is
unrealistic since it leads to an unacceptably large stop contribution
to the $K_{S}-K_{L}$ mass difference unless the top squark mass is around
1 $TeV$. Therefore, we are left with the second possibility which we will
examine in detail. From now on we assume that
${\tilde V_{td}}\simeq V_{td}/\sin\theta_{C},\;m_{\tilde W}\simeq 100GeV,\;
m_{LR}\simeq m_{\tilde q}$, and will study the
behavior of the lower bound on $\delta$ as a
function of the remaining free parameters - $z , \zeta , \tan\beta$, and
$m_{\tilde q}$.

Fig.4 displays a typical picture showing inconsistency
of the model.{\footnote{The displayed bound on the NEDM was
calculated for the gluino mass in the range $300-500\;GeV$ using
the standard formulas [16].}}
The lower bound exceeds the upper bound by one or two orders
of magnitude. Moreover,
the region allowed by the CP-asymmetry in $B\rightarrow \psi K_{S}$
is restricted to
the left upper corner of the plot.
 The reason for that can be easily understood. Indeed,
if the squarks are heavy,
the magnitude of the susy contribution is negligible as compared to that
of the CP-conserving SM box and sufficient CP-violation cannot be
produced regardless of how large the phases are.

Let us now consider the effect of each of the variable parameters.{\footnote{
We are assuming that $\delta$ belongs to the second
quarter. Even stricter lower bounds on $\sin\delta$ can be obtained
for $\delta$ in the second quarter due to a partial cancellation
in the numerator of Eq.(11).}}\\ \ \\
1.$\; \zeta-dependence\;(Fig.5).$\\
The lower bound on $\delta$ relaxes as we introduce the super-GIM
cancellation. This occurs due to the increasing share of the
CP-violating diagram in Fig.2b. However, theoretically one expects
 $\zeta$ to be of order unity due a large difference between the
stop and other squarks masses.\\ \ \\
2.$\; \tan\beta-dependence\;(Fig.6).$\\ 
The gap between the lower and upper bounds widens drastically as
$\tan\beta$ increases. Recalling that $h_{u}={gm_{u}\over {\sqrt {2}}
m_{W} \sin\beta}$ and $h_{d}={gm_{d}\over {\sqrt {2}}m_{W} \cos\beta}$,
it is easy to see that for large $\tan\beta$ the NEDM constraint becomes
stricter due to the large $h_{d}$ whereas the CP-violating contributions
to $B-\bar B$ mixing, proportional to powers of $h_{t}$, decrease.
We
do not consider the case $\tan\beta <1$ because of the
SUGRA constraints and the breakdown of perturbation theory
in this region. \\ \ \\
3.$\; z-dependence\;(Fig.7).$\\ 
Apparently, the incorporation of a partial cancellation ($z>0$) among
the CP-violationg contributions makes the lower bound rise.
One expects the natural value for $z$ to be around 1/2.

In all of these cases the regions allowed by the NEDM and $B\rightarrow
\psi K_{S}$ are at least
an order of magnitude apart.{\footnote{Note also that
small CP-phases are disfavored by the bound on the lightest Higgs
mass [8].}}
Furthermore, heavy squarks ($\geq 400\;GeV$)
are prohibited by large CP-asymmetries observed in $B\rightarrow \psi K_{S}$.
This condition is quite restrictive and may alone be sufficient
to rule out the model in the near future
(even if large CP-phases were allowed).

It should be mentioned that, in the limit of large $\tan\beta$, the
CP-violating neutralino and Higgs contributions to $B-\bar B$ mixing
become more important. However, the CP-phases are severely constrained
in this case ($\sim 10^{-3}$). Therefore, these
contributions do not lead to any considerable modifications of our analysis.

\section{Further Discussion}
The model under consideration has further implications for B-physics.
For instance, in this model, direct CP-violation is greatly suppressed in all
tree-level allowed processes as compared to what one expects in
the Standard Model. This provides another signature testable in the
near future.

The other angles of the unitarity triangle, $\alpha$ and $\gamma$,
can be determined in a similar manner from,
for example, $B_{d}\rightarrow \pi^{+}\pi^{-}$
and $B_{s}\rightarrow \rho K_{s}$ decays [11]. However,
in the Standard Model, one cannot determine these angles from the
decay rate evolution and relations
analogous to (5) precisely enough due
to considerable penguin contributions. To eliminate their influence,
one can use isospin and $SU(3)$ relations [18].
For instance, in order to determine $\alpha$, one needs to know
the rates of $B_{d}\rightarrow \pi^{+}\pi^{-}$,
$B_{d}\rightarrow \pi^{0}\pi^{0}$, and $B_{d}^{+}\rightarrow \pi^{+}\pi^{0}$
along with the rates of their charge conjugated processes. Then $\alpha$
can be found from certain triangle relations among the corresponding
amplitudes. In our case, however, this analysis becomes trivial due to
a vanishingly small interference between the tree and superpenguin
contributions. As a result, $\alpha$ can
be read off directly from the analog of Eq.(12).
The angles extracted in such
a way normally do not exceed a few degrees (modulo $180^{0}$)
and do not form a triangle [10].

In our model, the angles of the unitarity triangle do not
have a process-independent meaning. Indeed, contrary to the Standard Model,
they cannot be extracted from direct CP-violating processes simply because
such processes are prohibited.
Moreover, these angles are not related to the sides of the unitarity triangle.

Another important consequence of the model is that the CKM matrix is real
and orthogonal.
This, of
course, is also true for general (nonsupersymmetric) two Higgs doublet models
with FCNC constraints, in which CP is broken spontaneously [19].
As a result, $|V_{ub}|,\;|V_{td}|$, and $| \sin\theta_{C} V_{cb}|$ must form
a flat triangle. Presently, such a triangle is experimentally allowed
provided new physics contributes significantly to $\Delta m_{B_{d}}$ [20].
To determine the status of these models, a more precise determination of
$|V_{ub}|$ and $|V_{td}|$ is necessary. Orthogonality relations among
other CKM entries are less suitable for probing this class of models
due to the small CKM-phases involved.

We have considered in detail the case of the NMSSM. The results,
however, remain valid for NMSSM-like models with an arbitrary number
of sterile superfields $\hat N$. Indeed, since the $\hat N$'s
do not interact with matter fields directly, an introduction of a sterile
superfield does not affect the way CP-phases enter the observables.
The CP-violating effects will still be described by the diagrams in
Figs.2 and 3. Therefore, the argument we used also applies
to this more general situation: the CP-phase in left-right
squark mixing can be constrained via the gluino contribution to the NEDM,
whereas the phase in the gaugino-higgsino mixing can be constrained via
that of the chargino (assuming no accidental cancellation).
In the same way, we find that the lower and upper bounds on the CP-phases
are incompatible.
A nontrivial extension of the model, in which
a cancellation among various contributions to the NEDM can be
well motivated,
is necessary to rectify this problem.
{\footnote{ The possibility of such a cancellation
in certain susy models has recently been considered by a few authors.
It is, however, unclear whether this cancellation can be made natural
(see [17] and references therein).}}

\section{Conclusions}
We have analyzed the CP-asymmetry in 
$B\rightarrow \psi K_{s}$ decay within minimal supersymmetric models
with spontaneous CP-violation.
We have found that the
CP-asymmetry required by  $\sin 2\beta \geq 0.4$ can be accommodated
in these models only if the following conditions are met:\\
1. left-right squark mixing is maximal, $m_{LR}\simeq m_{\tilde q}$,\\
2. super-CKM mixing is enhanced,
   $\tilde V_{td} \simeq V_{td}/\sin\theta_{C}$,\\
3. the chargino is relatively light, $m_{\tilde W} \simeq 100\;GeV$,\\
4. the t-squark is lighter than $350-400\;GeV$.

Even if this is the case, the required CP-violating phases are
larger than those
allowed by the bound on the NEDM by one or two orders of magnitude.

We conclude that the model under consideration cannot accommodate
a large CP-asymmetry in $B\rightarrow \psi K_{s}$ while complying
with the bound on the NEDM.
Thus, if the Standard Model prediction gets confirmed, the model will
be recognized unrealistic.
  The result holds
true for models with an arbitrary number of sterile superfields.
To reconcile theory with experiment one would have to resort to essentially
nonminimal scenarios in which large CP-violating
phases are naturally allowed.

Note also that since, in this approach, CP-violation is
a purely supersymmetric effect, the model requires
the existence of a relatively light t-squark. This may conflict
with the Tevatron constraints on supersymmetry in the near future
(see, for example, [13]).

We have also discussed other testable predictions of the model.
Among them are the suppression of direct CP-violation and
orthogonality of the CKM matrix.

The author is grateful to L. N. Chang and T. Takeuchi for discussions
and reading of the manuscript, and to T. Falk for useful comments.

\newpage
{\bf Figure Captions}\\ \ \\
Fig. 1a,b $\;$ Most important contributions to $B-\bar B$ mixing. \\ \ \\
Fig. 2a,b $\;$ Dominant CP-violating contributions
           to $B-\bar B$ mixing (all possible permutations are implied).\\ \ \\
Fig. 3a,b $\;$  Contributions to the NEDM allowing to constrain the
           wino-higgsino and squark left-right mixing phases.\\ \ \\
Fig. 4    $\;\;\;$ Regions allowed by the CP-asymmetry in
          $B\rightarrow \psi K_{s}$ (upper) and the bound on the NEDM (lower).
          Typically, the region allowed by $B\rightarrow \psi K_{s}$ is
          much smaller than shown here due to the partial
          cancellation (see Fig.7, $z=1/2$). \\ \ \\
Fig. 5    $\;\;$ Lower bound on $\sin\delta$ 
          as a function of the super-GIM cancellation parameter $\zeta$:
          1 - $\;\zeta =1$,$\;$ 2 - $\;\zeta =1/2$,$\;$ 3 - $\;\zeta= 1/4$.
          \\ \ \\
Fig. 6    $\;\;$ Lower  bound on
          $\sin\delta$ as a function of $\tan\beta$:
          1 - $\;\tan\beta =1$, $\;$2 - $\;\tan\beta =25$.
          \\ \ \\
Fig. 7    $\;\;$ Lower bound on $\sin\delta$   
          as a function of the partial cancellation parameter $z$:
          1 - $\;z =0$,$\;$ 2 - $\;z =1/4$, $\;$3 - $\;z= 1/2$. 

\end{document}